# Ultra-low threshold pH sensor based on a whispering gallery mode microbubble resonator

YUE CAO,[1,2] JIN DAI, [1,2] XUBIAO PENG, [1,2] JIYANG MA, [1,2] AND QING ZHAO[1,2]

[1]Center for Quantum Technology Research and Key Laboratory of Advanced Optoelectronic Quantum Architecture and Measurements (MOE), School of Physics, Beijing Institute of Technology, Beijing 100081, China

[2]Beijing Academy of Quantum Information Sciences, Beijing 100193, China

**Abstract**：Laser sensing has a wide range of applications. In this paper, we propose a pH sensing laser with an ultra-low threshold and low sample consumption based on a whispering-gallery-mode microbubble resonator. Rhodamine 6G aqueous solutions with different pH values are used as the lasing gain media, which are injected through the microfluidic channel and interact with the high-quality-factor microbubble resonator to achieve lasing. Subtle pH changes of the aqueous solution lead to changes in lasing intensity in real time and the threshold reaches a minimum of 0.091 μJ/mm$^2$. The low pump energy density effectively avoids the self-aggregation and photobleaching effects of dye molecules present in high-concentration rhodamine 6G solutions. The lasing characteristics under different pH conditions were determined experimentally and theoretically, and the results are in good agreement. Due to the deprotonation of amino groups in highly alkaline environments, the lasing threshold is highly dependent on the pH of rhodamine 6G aqueous solutions. In the pH range of 10.16–13.14, the lasing intensity changes considerably with the increasing pH. The proposed pH-sensing laser exhibits a fast response time, low toxicity, and a high signal-to-noise ratio, making it promising for highly sensitive alkaline detection in biological applications.

## 1. Introduction

pH plays an important role in environmental, industrial, agricultural, and biomedical fields, and pH monitoring has important scientific practical applications[1-14]. Therefore, numerous pH sensing applications and technologies have been developed. Currently, fluorescence-based sensing applications are widely used[15, 16]; however, organic fluorescent dyes often have limitations such as photobleaching, phototoxicity, and background fluorescence interference, which limit the stability and accuracy of fluorescence-based sensing. Lasing exhibits excellent physical properties, such as high brightness, directionality, monochromaticity, and coherence[17, 18]; therefore, it has a wider range of applications than fluorescence. Whispering-gallery-mode microbubble resonators, which have natural reusable microfluidic channels[19-23], are indispensable in biochemical sensing. However, most lasers are large, which limits their potential applications in sensing. Microcavities with micrometer dimensions can be used for sensing trace samples. Photons confined in whispering-gallery-mode cavities interact with matter multiple times. Therefore, whispering-gallery-mode optical microcavities have an extremely low threshold and high sensitivity, making them ideal platforms for optical sensing. At present, whispering-gallery-mode microcavity sensing has numerous applications. For example, Wang et al. developed a pH sensor based on liquid crystal droplets by doping PBA(4'-pentyl-biphenyl-4-carboxylic acid) into 5CB(4-cyano-4'-pentylbiphenyl) and verified its feasibility as a penicillin enzyme activity monitor[24]。Fan et al. proposed a covert optical cryptographic protocol based on silk protein microlaser arrays, applying whispering-gallery-mode microcavities to optical anti-counterfeiting[25]. Ruan et al. unified the manufacturing and application of laser textiles, paving the way for the manufacturing of novel wearable laser devices[26]. A microbubble resonator is a type of whispering-gallery-mode resonator, which is composed of a silica microbubble with a diameter of approximately 100 μm and a wall thickness of approximately 1 μm. The microfluidic channel formed by the hollow structure facilitates the injection of different types of gain medium solutions.

Rhodamine dyes have been widely studied and applied due to their excellent stability and photostability [27]. Among these dyes, rhodamine 6G is characterized by excellent water solubility, high quantum yield, and low cost. The fluorescence of rhodamine 6G changes when the pH around its molecule changes, which is attributed to the protonation and deprotonation of amino groups. As reported previous studies, when the concentration increases, the quantum yield decreases due to the self-assembly of rhodamine 6G dye molecules in aqueous solution, leading to a higher lasing threshold[28-30]. To achieve better sensitivity and eliminate the impact of photobleaching under low-lasing-threshold conditions, it is important to select a suitable concentration of rhodamine 6G solution as the gain medium, which is challenge. The ultra-high-quality factor of the whispering-gallery-mode microcavity can compensate for this disadvantage of rhodamine 6G. In this study, we selected an ultra-low concentration of rhodamine 6G solution (0.1 mM) to prevent the self-aggregation of dye molecules and reduce pollution.

In this paper, we propose a low-threshold microfluidic pH-sensing laser. We selected an ultra-low concentration of rhodamine 6G aqueous solution as the gain medium and combined it with a high-quality-factor whispering-gallery-mode microbubble cavity. Due to the high quality factor of the cavity(up to 108), the unique evanescent light field inside the microbubble promotes the interaction between the microresonator and gain medium. Therefore, even at low concentrations of rhodamine 6G solution, the threshold reaches a minimum of 0.091 µJ/mm$^2$. At this threshold, rhodamine 6G molecules are effectively protected from pump light damage, extending their lifetime. In addition, the volume of the resonant cavity is only approximately 550 pL, and the sample can be reused by changing the pressure. Therefore the proposed laser can be applied to pH sensing of small sample solutions.

2. **Experimental setup**

The experimental setup is presented in Figure 1(a). The microbubble is pumped by an optical parametric oscillator (OPO) laser at 532 nm with a pulse width of 5 ns and a repetition of 10 Hz. The energy of the pumped light is controlled by a polarizer. The spot diameter is adjusted by a beam expander to fully illuminate the microbubble resonator. The lasing generated by the excitation of rhodamine 6G dye molecules is collected by a 20x objective lens and filtered out by dichroic mirrors and a long-pass filter. Finally, an optical fiber is used to collect the laser beam and transmit it to a spectral analyzer (Ocean Optics, Maya2000 Pro, 1-nm resolution; Teledyne Princeton Instruments, SpectraPro HRS-750, 0.05-nm resolution).

The fabrication process of the microbubble resonator is as follows. First, piranha solution is heated to 155 °C to remove the polymer layer on the surface of the capillary tube (Polymicro Technologies TSP100170, outer diameter 140 µm, inner diameter 100 µm, wall thickness 20 µm). The capillary tube is etched with 5% hydrofluoric acid to an outer diameter of approximately 125 µm. Then, the capillary tube is stretched to a center outer diameter of approximately 35 µm while being heated with a hydrogen flame. Thereafter, one end of the capillary tube is sealed with ultraviolet glue, while the other end is pressurized with an injection pump. In addition, a carbon dioxide laser is used to irradiate the thinnest part of the capillary tube, causing it to melt and expand to form a microbubble with an outer diameter of approximately 100 μm and a wall thickness of approximately 1 μm, as displayed in Figure 1(b)[31]. The simulation of the optical mode distribution in the microbubble resonator filled with rhodamine 6G solution is illustrated in Figure 1(c). Due to the ultra-thin wall thickness of the resonator, its mode exhibits a strong evanescent field, which effectively interacts with rhodamine 6G to generate laser.

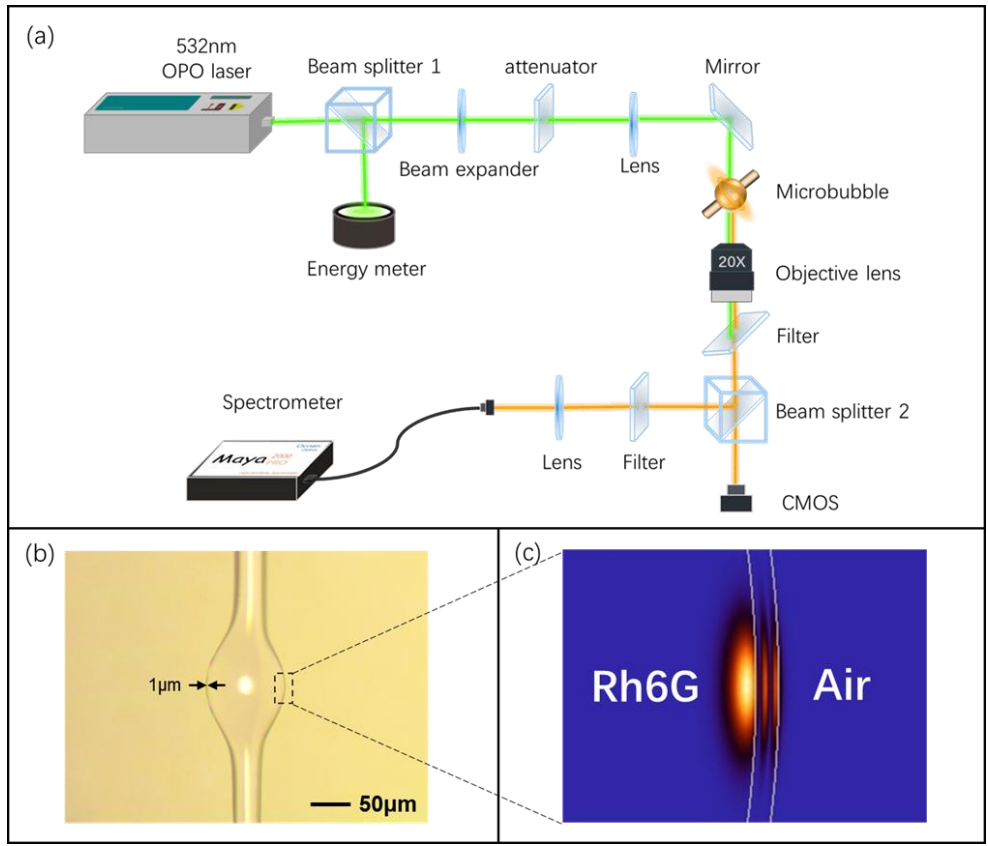

Fig. 1 (a) Schematic diagram of the experimental setup. Optical parametric oscillator (OPO) laser with a pulse width of 5 ns, repetition frequency of 10 Hz, and output wavelength of 532 nm. Beam splitter 1 has a 1/1 ratio, while beam splitter 2 has a 1/10 ratio. The diameter of the microbubble is 100 μm, and the wall thickness is approximately 3 μm. (b) Microscopic images of the microbubble resonator. (c) Simulation of the rhodamine 6G laser mode in the microbubble resonator.

In the experiment, the concentration of rhodamine 6G dye was set to 0.1 mM. The pH of rhodamine 6G aqueous solution was adjusted by adding NaOH solution to DI water, and the pH of the gain medium was calibrated using a calibrated pH meter.

## 3. Results and discussion

We verified the feasibility of laser generation in the microbubble resonator using rhodamine 6G aqueous solution (pH = 6.89) as the gain medium. First, we observed changes in the lasing spectrum with increasing pump energy density using a low-resolution spectrometer.

As illustrated in Figure 2(a), when the pump energy is low (0.073 µJ/mm$^2$), a relatively wide fluorescence peak can be observed. When the pump energy increases to 0.161 µJ/mm$^2$, a clear and sharp lasing peak can be observed above the fluorescence peak. As the energy continues to increase, both the fluorescence and lasing peaks are enhanced; however, the changes in the lasing peaks are more pronounced. We integrate the lasing energy as a function of pump energy density, as illustrated in Figure 2(b). The fitting results indicate that when pH = 6.89, the threshold reaches a minimum of 0.091 µJ/mm$^2$.

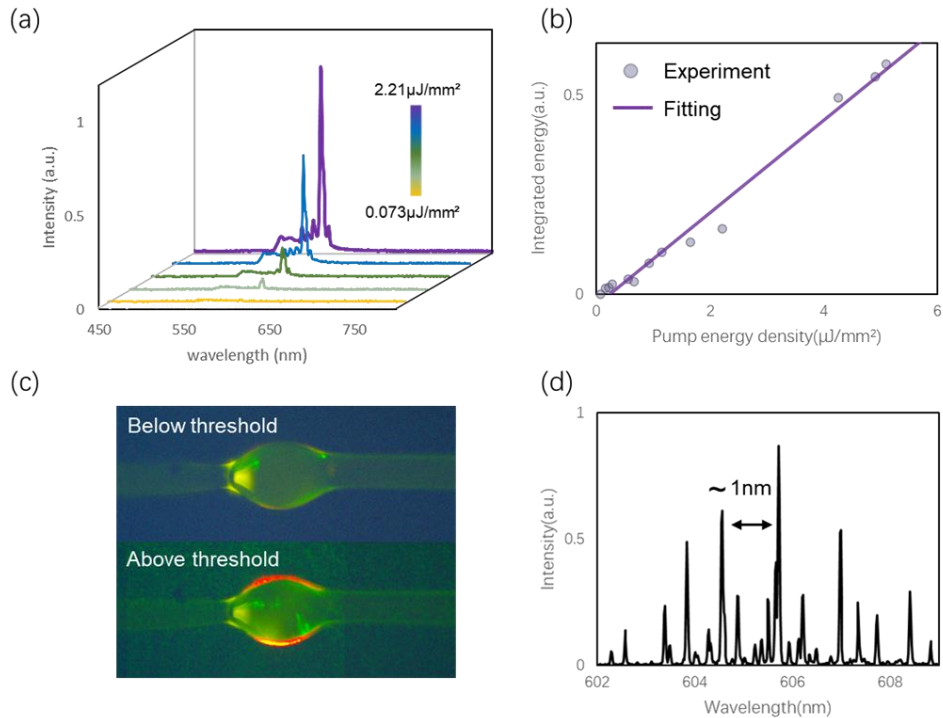

Fig. 2 (a) Comparison of spectrum above and below the threshold collected using a low-resolution grating spectral analyzer. (b) Relationship between laser output intensity and pump energy. Dots represent experiment data, while straight lines represent fitting values. (c) Image of microbubble resonator with the pump energy density above and below the threshold. The edges of the microbubble are characterized by weak fluorescence emission below the threshold and strong orange lasing emission above the threshold. (d) High-resolution lasing spectrum collected by a

1200-g/mm grating spectral analyzer with a pump energy density of 2.0 µJ/mm².

Figure 2(c) presents images of the microbubble cavity before and after lasing occurs. It can be seen that when the pump energy density is below the threshold, the microbubble resonator emits uniform yellow fluorescence. When the pump energy is above the threshold, bright orange lasing occurs around the equator of the microbubble resonator, which is consistent with the characteristics of whispering-gallery-mode lasing. Figure 2(d) displays the high-resolution lasing spectrum measured using a 1200-g/mm grating spectral analyzer at a pump energy density of 2.0 µJ/mm². The distance between adjacent lasing peaks is approximately 1 nm, which is consistent with the theoretical distance between lasing peaks in whispering gallery mode: $FSR = \lambda^2/(2\pi n r)$, where $\lambda$ (= 605.5 nm) is the center wavelength, n (= 1.45) is the refractive index of the silica fiber, and r (= 50 µm) is the radius of the microbubble cavity.

In summary, the use of rhodamine 6G aqueous solution (pH = 6.89) as the gain medium can generate whispering-gallery-mode lasing in a microbubble resonator, and the threshold reaches a minimum of 0.091 µJ/mm².

To further explore the feasibility of implementing a pH-sensing laser using rhodamine 6G aqueous solution as the gain medium, we investigated the relationship between the lasing intensity and pH while keeping the pump energy constant. As illustrated in Figure 3(a), when the pump energy is maintained at approximately 5.0 µJ/mm², the lasing intensity gradually decreases with an increase in the pH of the rhodamine 6G aqueous solution. When the pH of the rhodamine 6G aqueous solution reaches 13.59, the pump energy density falls below the threshold, resulting in the quenching of lasing.

To examine the influence of pH changes on lasing intensity more intuitively, we employed a higher-resolution spectral analyzer to explore pH regions with sharp changes. The spectra measured with a 300-g/mm grating spectral analyzer are presented in Figure 3(b). It can be seen that in the 10.16–12.15 region, as the pH increases, the emission peak decreases and the central wavelength shifts considerably.

Next, we investigated the variation in lasing intensity with pump energy density when rhodamine 6G solutions with different pH levels were used as the gain medium. As illustrated in Figure 3(c), when the pump energy density exceeds the threshold, the lasing intensity increases linearly with the increase in pump energy density. When the pH increases to 13.56, only weak lasing can be observed even when the pump energy exceeds 25.978 µJ/mm². Figure 3(d) demonstrates that as the pH increases, the lasing threshold increases nonlinearly, with the changes becoming more pronounced when the pH is above 9.08. As illustrated by the colors of solutions with different pH levels (displayed in the inset of Figure 3(d)), the solution appears bright yellow when the pH is low and changes to a light pink color as the pH increases.

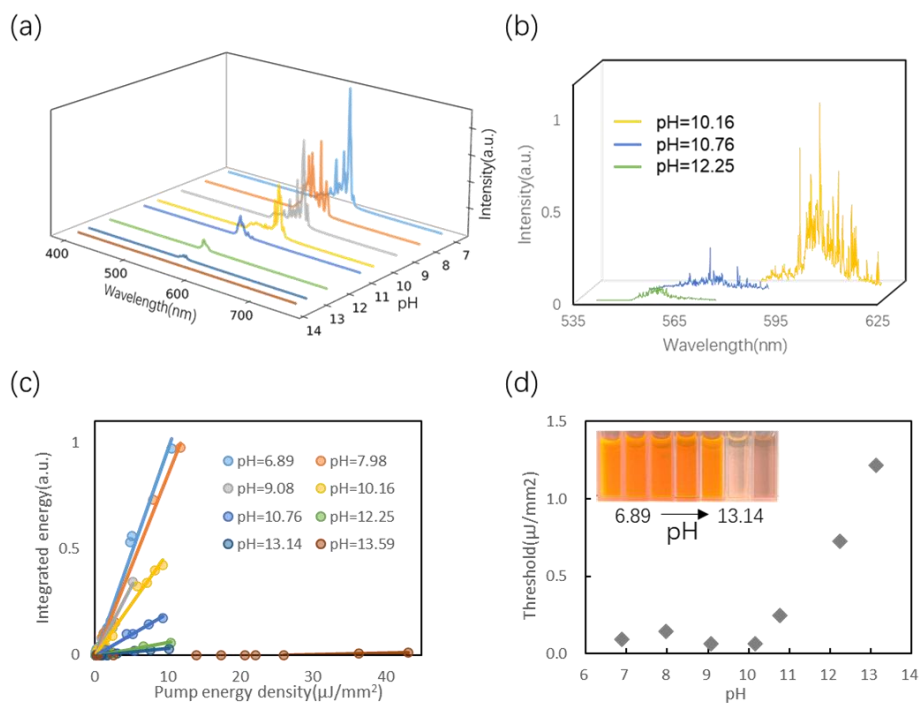

Fig. 3 (a) Low-resolution emission spectrum of rhodamine 6G aqueous solution with different pH levels (pump energy maintained at approximately 5.0 µJ/mm²). The lasing intensity gradually decreases with the increase in pH. When the pH reaches 13.59, lasing lasing is quenched. (b) High-resolution emission spectrum in the pH range of 10.16–12.25. (c) Integrated lasing emission intensity as a function of pump energy density under different pH conditions. Solid lines represent linear fits above the threshold. The lasing thresholds are approximately 0.091 µJ/mm² (pH = 6.89), 0.141 µJ/ mm² (pH = 7.98), 0.062 µJ/ mm² (pH = 9.08), 0.066 µJ/ mm² (pH = 10.16), 0.249 µJ/ mm² (pH = 10.76), 0.728 µJ/ mm² (pH = 12.25), 1.218 µJ/ mm² (pH = 13.14), and 25.978 µJ / mm² (pH = 13.59). (d) Changes in the lasing threshold with increasing pH. Inset: Changes in the color of rhodamine 6G aqueous solution with different pH levels.

To gain a deeper understanding of the performance change of the rhodamine 6G microbubble laser under different pH conditions, we performed theoretical calculations.

According to relevant theories, the threshold of microbubble cavities is given by the following equation:

$$I_{th} = \frac{\gamma}{1-\gamma}, \quad (1)$$

where $\gamma$ represents the proportion of gain molecules in the excited state at the threshold and can be expressed as follows:

$$\gamma = \frac{\sigma_a(\lambda)}{\sigma_a(\lambda) + \sigma_e(\lambda)}[1 + \frac{Q_{abs}}{Q_0}], \quad (2)$$

where $\sigma_a(\lambda)$ and $\sigma_e(\lambda)$ represent the absorption cross section and emission cross section of the dye, respectively. $\sigma_a(\lambda) = A(\lambda)ln10/n_t l$, where $A(\lambda)$ denotes the absorbance, $n_t$ denotes the total concentration of the dye, and $l$ denotes the length of the optical path. $\sigma_e = \Phi g(\lambda)\lambda^4/8\pi c\tau n_2$, where $n_1$ is the refractive index of the microbubble resonator, $n_2$ is the refractive index of the rhodamine 6G aqueous solution, and $\Phi$ and $\tau$ represent the quantum yield and fluorescence lifetime of rhodamine 6G, respectively. In addition, $g(\lambda) = I(\lambda)/\int I(\lambda)d\lambda$ is the normalization function of the fluorescence spectrum, $Q_0$ is the quality factor of the microbubble resonator, $Q_{abs} = 2\pi m/\eta\lambda n_t\sigma_a(\lambda)$ is the quality factor related to dye absorption, and $\eta$ is the proportion of the mode volume in the region where the gain medium is located to the total echo wall mode volume. According to the simulation results, $\eta$ is approximately 0.76. $m = n_1/n_2$ is the effective refractive index.

It can be seen that changes to any parameter in equation (2) will affect the laser threshold of the microbubble cavity. We believe that the pH dependence displayed in Figure 3 is mainly due to changes in the gain medium; specifically, the decrease in the luminescence of rhodamine 6G in highly alkaline environments is due to the deprotonation of amino groups. To further investigate the effect of dye parameters on the laser threshold, we measured the corresponding changes in the dye absorption cross section, quantum yield, and fluorescence lifetime, as illustrated in Figure 4(a).

When the pH is below 10.16, the performance of the dye is almost unaffected, and the laser threshold fluctuates only slightly. However, when the pH is above 10.16, there is a considerable decrease in the absorption cross section, fluorescence lifetime, and quantum yield. Notably, the rate of change in quantum yield is much higher than that of the absorption cross section and fluorescence lifetime. When the pH of the solution increases from 10.16 to 13.14, the absorption cross section decreases from 8.32 × 10⁻¹⁸ cm2 to 2.17 × 10⁻¹⁸ cm2, and the fluorescence lifetime decreases from 5.38 ns to 2.35 ns (a 3.85- and 2.33-fold decrease, respectively). The quantum yield exhibits a stronger pH dependence in this range, achieving a 6.85-fold decrease from 45.45% to 6.64% decreasing from 45.45% to 6.64%, reaching as low as 0.146 times the original value. Therefore, in this pH range, the laser threshold increases considerably. The light blue section in Figure 4 displays significant changes in the dye absorption cross section, quantum yield, and fluorescence lifetime between pH values of 10.16 and 13.14.

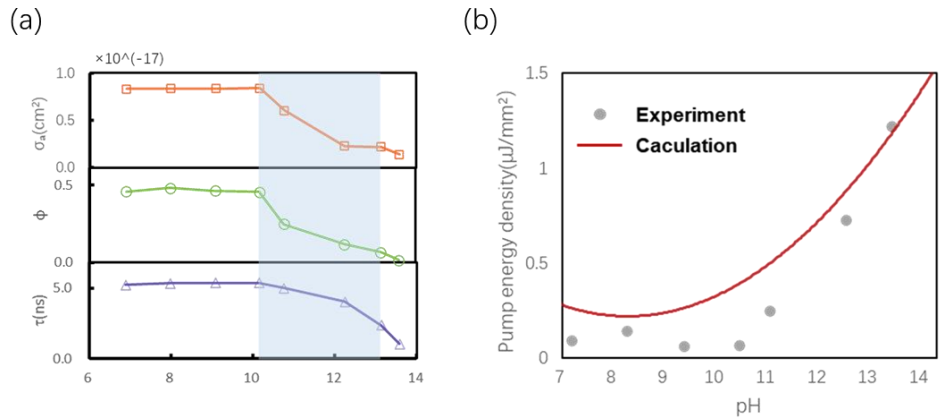

Fig. 4 (a) Functional relationship between the properties of rhodamine 6G and pH. The properties include the absorption cross section ($\sigma_a$), quantum efficiency ($\Phi$), and excited-state lifetime ($\tau$). (b) Comparison of normalized lasing thresholds determined experimentally and theoretically. Gray dots represent the experimental results, while the red solid line represents the theoretical calculation results.

As illustrated in Figure 4(b), there is good agreement between the theoretically calculated and experimentally measured thresholds as the pH of the solution increases. These results indicate that a microbubble resonator with rhodamine 6G as the gain medium can be used as a pH detector for trace samples. To further validate this conclusion, we conducted a control experiment to evaluate whether the sensing response of the pH sensor changes under different salt concentrations.

Figure 5(a) displays the variation in emission intensity (pump energy density ≈ 4.0 µJ/mm²) with an increasing NaCl concentration at a fixed pH (pH = 6.51). In the salt concentration range of 0.01–0.8 M, there is no significant change in laser intensity. We also measured the emission spectrum at different concentrations (Figure 5(b)), which indicated only slight changes in the central wavelength and waveform. These results suggest that the effect of salt concentration on the sensing response of pH sensors can be ignored; that is, it is OH−, not Na+, that causes the change in dye laser performance.

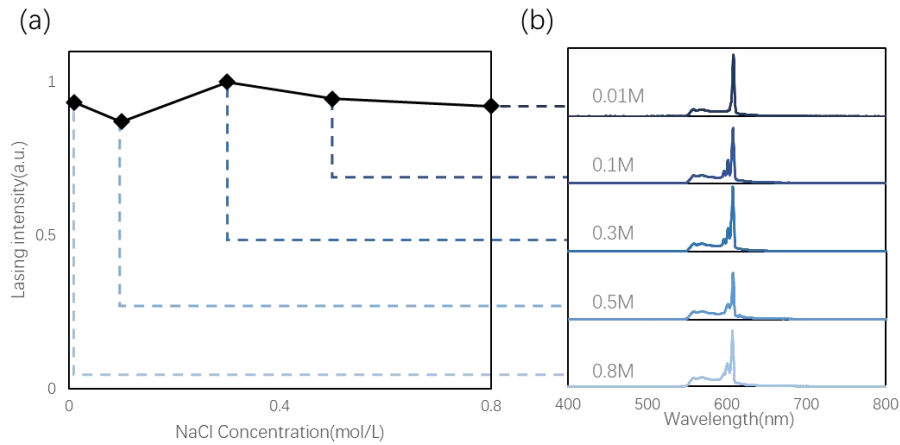

Figure 5 (a) Integrated laser emission intensity as a function of pump energy density under different NaCl concentrations. (b) Lasing emission spectrum at NaCl concentrations of 0.01, 0.1, 0.3, 0.5, and 0.8 M at a pump energy of 4 µJ/mm².

## 4. Conclusion

In this study, we developed a pH-sensing laser based on a whispering-gallery-mode microbubble resonator. We used a high-quality-factor microbubble resonator, allowing the experiment to be conducted at a low pump energy density. When the pH of rhodamine 6G aqueous solution is 6.89, the threshold reaches a minimum of 0.091 µJ/mm². By further adjusting the solution pH, significant changes in lasing intensity are observed, demonstrating the high dependence of the laser intensity and threshold on pH. Subsequently, we measured the influencing factors leading to changes in laser intensity, including the absorption cross section, fluorescence lifetime, and quantum yield. The results indicate that in the solution pH range of 10.16 to 13.14, these influencing factors all exhibit significant changes, consistent with the changes in laser intensity. The sensor proposed in this study can be applied to ultra-low and trace pH sensing. In future work, we plan to use previously studied fluorescent proteins for sensing, explore the application of lasers to biosensing, and investigate the possibility of in vivo laser sensing.

**Disclosures.** The authors declare no conflicts of interest

**Data availability.** Data underlying the results presented in this paper are not publicly available at this time but may be obtained from the authors upon reasonable request.